# Thermodynamic Driving Forces for Substrate Atom Extraction by Adsorption of Strong Electron Acceptor Molecules

Paul Ryan[1,2], Philip James Blowey[1,3], Billal S. Sohail[4], Luke A. Rochford[1], David A. Duncan[1], Tien-Lin Lee[1], Peter Starrs [1,5], Giovanni Costantini[4], Reinhard J. Maurer[4*], David Phillip Woodruff[3*]

*(1) Diamond Light Source, Harwell Science and Innovation Campus, Didcot, OX11 0DE, UK*
*(2) Department of Materials, Imperial College, London SW7 2AZ, UK*
*(3) Department of Physics, University of Warwick, Coventry CV4 7AL, UK*
*(4) Department of Chemistry, University of Warwick, Coventry CV4 7AL, UK*
*(5) School of Chemistry, University of St. Andrews, St. Andrews, KY16 9AJ, UK*

**Keywords**: surface structure; TCNQ; STM; NIXSW

## Abstract

A quantitative structural investigation is reported, aimed at resolving the issue of whether substrate adatoms are incorporated into the monolayers formed by strong molecular electron acceptors deposited onto metallic electrodes. A combination of normal-incidence X-ray standing waves, low energy electron diffraction, scanning tunnelling microscopy and X-ray photoelectron spectroscopy measurements demonstrate that the systems TCNQ and $F_4$TCNQ on Ag(100) lie at the boundary between these two possibilities and thus represent ideal model systems with which to study this effect. A room-temperature commensurate phase of adsorbed TCNQ is found not to involve Ag adatoms, but to adopt an inverted bowl configuration, long predicted but not previously identified experimentally. By contrast, a similar phase of adsorbed $F_4$TCNQ does lead to Ag adatom incorporation in the overlayer, the cyano endgroups of the molecule being twisted relative to the planar quinoid ring. Density functional theory (DFT) calculations show that this behaviour is consistent with the adsorption energetics. Annealing of the commensurate TCNQ overlayer phase leads to an incommensurate phase that does appear to incorporate Ag adatoms. Our results indicate that the inclusion (or exclusion) of metal atoms into the organic monolayers is the result of both thermodynamic and kinetic factors.

---

* corresponding authors d.p.woodruff@warwick.ac.uk, r.maurer@warwick.ac.uk

**INTRODUCTION**

The electronic properties of devices based on organic semiconductors are strongly influenced by the role of metal-organic interfaces at conductive electrodes, motivating a significant number of surface science studies of related model systems. One electron acceptor organic molecule of particular interest is 7,7,8,8-tetracyanoquinodimethane (TCNQ) which, together with its more strongly electron accepting fully fluorinated variant, $F_4$TCNQ, is frequently used as a molecular dopant in organic devices and for work function engineering [1-3]. As such there have been several studies of TCNQ adsorption on coinage metal surfaces, particularly with a (111) orientation (*e.g.* [4-13]). Spectroscopic studies clearly demonstrate charge transfer from the metal surface, leading to re-hybridisation of the intra-molecular bonding that relaxes the rigidity of the planar gas-phase molecule. The results of many density functional theory (DFT) calculations have predicted that the adsorbed molecule adopts an inverted bowl or umbrella conformation, with the cyano N atoms bonding to the surface while the central quinoid ring is up to 1.4 Å higher above the surface. However, prior to the work reported here, there was no published quantitative experimental structural evidence to demonstrate the existence of this adsorption geometry.

In contrast to the implicit assumption in these earlier studies that the molecular electron acceptors form a purely organic layer, there are two cases, namely $F_4$TCNQ on Au(111) [14] ((P. Mousley *et al*, unpublished results)) and TCNQ on Ag(111)[15], in which adsorption has been unequivocally shown to lead to incorporation of metal adatoms to form two-dimensional metal-organic frameworks (2D-MOFs) on the surface. This structural modification can cause significant changes in the surface dipoles and thus in the electronic structure of the interface, so understanding (and ideally being able to predict) the conditions that lead to this effect have a wide relevance. To gain a better understanding of this phenomenon, in this work we have investigated the adsorption of both TCNQ and the significantly stronger electron acceptor molecule, $F_4$TCNQ, on Ag(100), a surface with a work function intermediate between that of the more noble Au and that of the more easily oxidised Cu. These specific "intermediate" model systems were chosen to explore the boundaries between the formation of pure organic and mixed metal-organic phases and achieve improved insight into the factors determining the nature of these thin molecular layers on metal electrodes.

Our general approach is the use of experimental quantitative structural studies complemented by dispersion-corrected DFT calculations. The experimental structural data are obtained from the use of the normal-incidence X-ray standing wave technique (NIXSW).[16] The general XSW technique [17] exploits the X-ray standing wave created by the interference of an incident X-ray wave and the resulting Bragg-reflected wave which shifts through the crystal as one scans through the Bragg condition. Monitoring the absorption of this standing wave in atoms of interest by measuring the element-specific X-ray fluorescence or core level photoemission allows one to determine the location of the absorbing atom relative to the Bragg planes from a wide range of materials.[16] Using core-level photoemission to monitor the absorption provides not only elemental specificity, but also chemical-state specificity, distinguishing the heights of C atoms in CH, CF, CN and CC bonds in adsorbed TCNQ and $F_4$TCNQ.

Using this approach, we recently showed that TCNQ adsorbed on Ag(111) does not adopt the inverted bowl configuration;[15] instead, the molecule is twisted on the surface, the cyano N atoms adopting two significantly different heights above the surface relative to the central planar quinoid ring. This geometry was found to be attributable to the presence of Ag adatoms in the overlayer to generate an Ag-TCNQ 2-D MOF. Previous evidence for similar incorporation of metal adatoms into the molecular layer had been suggested for $F_4$TCNQ on Au(111), based on STM imaging,[14] but only very recently has it been possible to unequivocally identify the presence and quantitatively determine the location of the Au adatoms in the resulting 2-D MOF, by means of surface X-ray diffraction (P. Mousley *et al*, unpublished results). Our objective now is to gain a better understanding of the conditions determining whether or not adsorption of an electron acceptor molecule leads to this type of surface reconstruction.

Here we report the results of NIXSW experiments and dispersion-inclusive DFT calculations for the commensurate adsorption phase of TCNQ on Ag(100) formed at room temperature that provides the first proven and quantitatively determined example of adsorption of TCNQ in an inverted bowl configuration at an unreconstructed surface. We further find that annealing of this low coverage phase of TCNQ to higher temperatures leads to an incommensurate phase that STM images suggest may well involve Ag adatoms. By switching to a stronger electron acceptor, namely $F_4$TCNQ, we identify a commensurate adsorption phase, formed without the need for annealing, for which NIXSW data and DFT calculations clearly indicate that the molecule has a twisted molecular conformation that can be attributed to the presence of Ag adatoms. Based on thermodynamic arguments, we construct a rationale for the formation of adatoms as a function of adsorbate strength, surface reactivity, and temperature.

## METHODS

### Experimental Methods

Experimental characterisation of the adsorption phases of TCNQ and $F_4$TCNQ on Ag(100) was performed using STM and low-current (microchannel plate) low energy electron diffraction (MCP-LEED) in a UHV surface science chamber at the University of Warwick, and by MCP-LEED and SXPS in the UHV end-station of beamline I09 of the Diamond Light Source. A well-ordered clean Ag(100) sample was cleaned *in situ* by cycles of 1 keV $Ar^+$ ion scattering and annealing in both chambers. Single molecular monolayer structures were prepared by vacuum deposition from evaporation sources installed in the chambers. NIXSW experimental data were collected from TCNQ and $F_4$TCNQ on Ag(100), by measuring the C 1s, N 1s, and F 1s photoelectron spectra as the incident photon energy was stepped through the (200) Bragg reflection very close to normal incidence to the (100) surface around a photon energy of 3036 eV. Comparisons of the relative intensity of the component peaks as a function of photon energy with standard formulae taking account of the backward forward asymmetry of the angular dependence of the photoemission allowed the optimum values of the coherent fraction and coherent positions to be determined [16].

### Computational Methods

The theoretical analysis has been performed at the Density Functional level of theory, using the all-electron numeric atomic orbital package – Fritz-Haber Institute *ab initio* molecular simulations package (FHI-aims).[26] To evaluate exchange and correlation, we use the generalised gradient approximation (GGA) variant by Perdew, Burke and Ernzerhofer (PBE)[18] coupled with dispersion correction schemes to account for significant van der Waals contributions to the total energy. The two dispersion schemes we utilised are conceptually different in approach with the Tkatchenko-Scheffler van-der-Waals surface correction (PBE+vdW$^{surf}$)[27] being a pairwise additive scheme, whereas the recently proposed non-local many-body dispersion (PBE+MBD-NL)[28] goes beyond this limit. As shown in the main text, despite the difference between these schemes we find close comparison between experimentally observed height parameters and theoretically derived values.

The adsorption structures were modelled as a periodically repeated cell comprising a single unit mesh described by experimentally determined matrices of the substrate lattice vectors, containing a single TCNQ or F$_4$TCNQ molecule. The Ag(100) surface was modelled as a slab consisting of four atomic layers and separated from its periodic image by a vacuum gap exceeding 90 Å. The coordinates of the atoms in the bottom two layers of the Ag slab were constrained to the bulk truncated structure of Ag and the positions of the adsorbate and top two layers of the substrate were allowed to relax. To minimize strain within the optimisation step, pre-calculated lattice constants were used for each dispersion scheme: PBE+vdW$^{surf}$ (4.14 Å) and PBE+MBD-NL (4.10 Å). In the case of PBE+vdW$^{surf}$, we exclude interactions between Ag atoms. The Brillouin zone was sampled with an 8×8×1 Monkhorst-Pack k-grid and the geometries were optimised to below a force threshold of 0.025 eV Å$^{-1}$. FHI-aims contains built-in basis sets for each atomic species regarding basis functions, integration grid and numerical accuracy of the Hartree potential. All equilibrium structures were optimised with the default "light" and then "tight" basis set. All calculation input and output files have been deposited as a data set in the NOMAD repository and are freely available at https://dx.doi.org/10.17172/NOMAD/2022.01.11-2.

**RESULTS AND DISCUSSION**

**Experimental surface characterisation**

Using the combination of STM and low energy electron diffraction (LEED), room temperature adsorption of both TCNQ and F$_4$TCNQ led to commensurate phases, $\begin{pmatrix} 3 & 1 \\ -2 & 3 \end{pmatrix}$ for TCNQ (previously reported in earlier STM [19] and LEED [20] studies where it is described by a unit mesh $\begin{pmatrix} 3 & 1 \\ 1 & 4 \end{pmatrix}$ with an acute included angle), and $\begin{pmatrix} 4 & 2 \\ -2 & 2 \end{pmatrix}$ for F$_4$TCNQ. STM images of these two phases are shown in Figure 1.

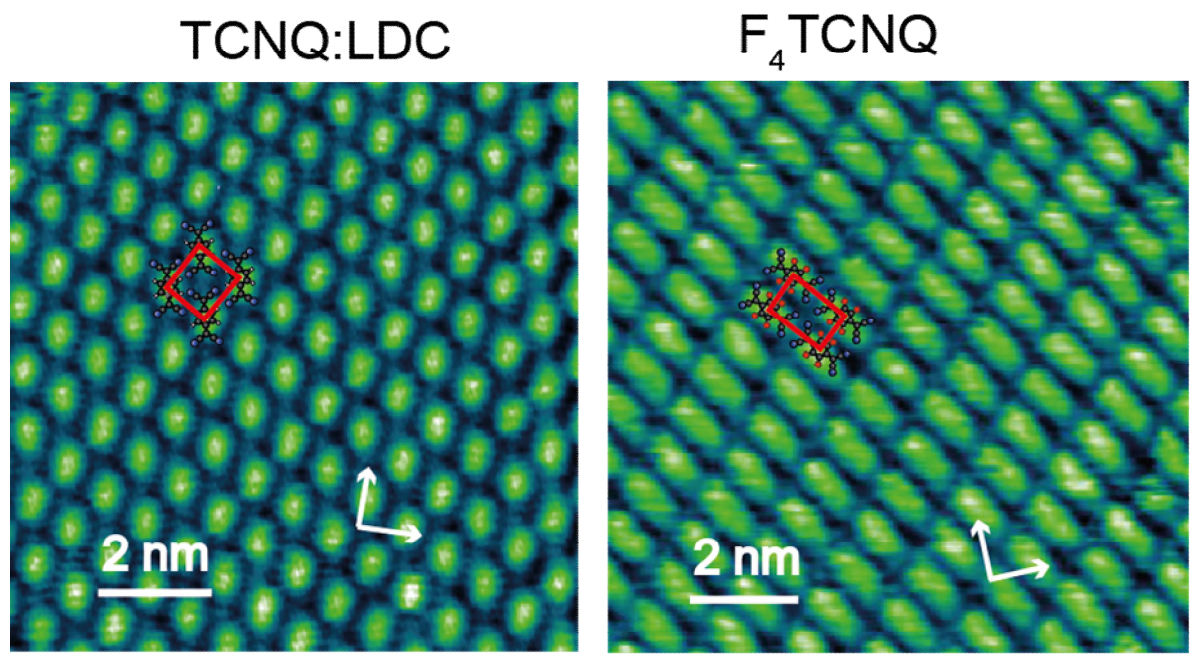

**Figure 1** STM images of the commensurate phases of TCNQ and F$_4$TCNQ on Ag(100). Superimposed on each image is the surface unit mesh (red) and a simplified schematic of the molecular structure assuming each of the elongated rectangular features, which are consistent in size and general appearance, can be attributed to molecules lying down on the surface. The arrows show <110> directions in the surface. STM tunnelling conditions (sample bias and tunnelling current): TCNQ:LDC, 0.25 V, 150 pA; F$_4$TCNQ, -1.00 V, 75 pA, chosen to optimise imaging contrast.

Because only commensurate phases can be fully modelled by DFT calculations, we concentrate in this paper on the properties of these two phases. However, we note that different deposition conditions of coverage or annealing led to four further ordered adsorption phases of TCNQ, but no other structures were observed for F$_4$TCNQ adsorption. STM images and LEED patterns, together with simulated LEED patterns obtained using the LEEDpat program,[21] are shown in Figure S1 for all of these phases. Their main characteristics and preparation conditions are summarised in Table S1. Notice that the molecular packing density of the commensurate F$_4$TCNQ phase is significantly lower than that of the commensurate phase of TCNQ on the same surface. The lowest coverage, commensurate $\begin{pmatrix} 3 & 1 \\ -2 & 3 \end{pmatrix}$ TCNQ phase we denote as TCNQ:LDC (low density commensurate). Two of the additional TCNQ phases we denote as higher-density incommensurate phases 1 and 2 (TCNQ:HDI1 and TCNQ:HDI2); these were formed by higher coverage exposure (TCNQ:HDI1) followed by subsequent annealing (TCNQ:HDI2). The TCNQ:HDI1 phase has also been reported previously [20], albeit assigned a slightly different matrix. Two further ordered phases, formed by heating the TCNQ:LDC phase to temperatures in the range up to ~340°C, were always found to coexist with one another and with the TCNQ:LDC phase and disordered regions. The associated STM images of these last two phases, included in Figure S1, show these structures to be based on the 'windmill' motif of four TCNQ molecules arranged like the four vanes of a windmill, a motif that has been reported in a number of studies of TCNQ and $F_4$TCNQ coadsorbed on coinage metal surfaces with transition (*e.g.*[7, 14, Error! Bookmark not defined., 22]) and alkali (*e.g.* [7, 12]) metal atoms. For this reason, we label these two 'windmill' phases TCNQ:W1 and TCNQ:W2. The TCNQ:W2 phase comprises entirely an ordered array of these windmill motifs, whereas the TCNQ:W1 phase also appears to contain some additional TCNQ molecules between groups of four windmill motifs. A significant feature of the TCNQ:W1 and TCNQ:W2 STM images, particularly for the TCNQ:W2 phase, is that the 'windmills' have bright centres that may indicate the presence of an Ag adatom at this location. The possible significance of this feature will be discussed later in this paper. Note that for all the phases other than TCNQ:W1 and TCNQ:W2, which always coexisted leading to a mixed-phase LEED pattern that was difficult to disentangle, obtaining a good fit to the LEED patterns with LEEDpat provided an accurate basis for determining the overlayer matrix, independent of any possible calibration errors or image drift in STM imaging.

The commensurate adsorption phases for TCNQ and $F_4$TCNQ on Ag(100) were also characterised by soft X-ray photoelectron spectroscopy (SXPS). C 1s spectra recorded from the TCNQ:LDC phase are compared with those for the ordered phase of $F_4$TCNQ in Figure 2, while the N 1s and F 1s spectra are shown in Figure S3 of the Supporting Information. These show a single N 1s peak, which indicates that all N atoms occupy closely similar chemical environments. The C atoms contributing to the different chemically shifted C 1s components in Figure 2 are labelled according to the inset schematic diagrams of the molecules. The relative binding energies of these components are consistent with electron transfer from the metal to the TCNQ molecule as reported in several previous studies (*e.g.*[12, 15, 23, 24]).

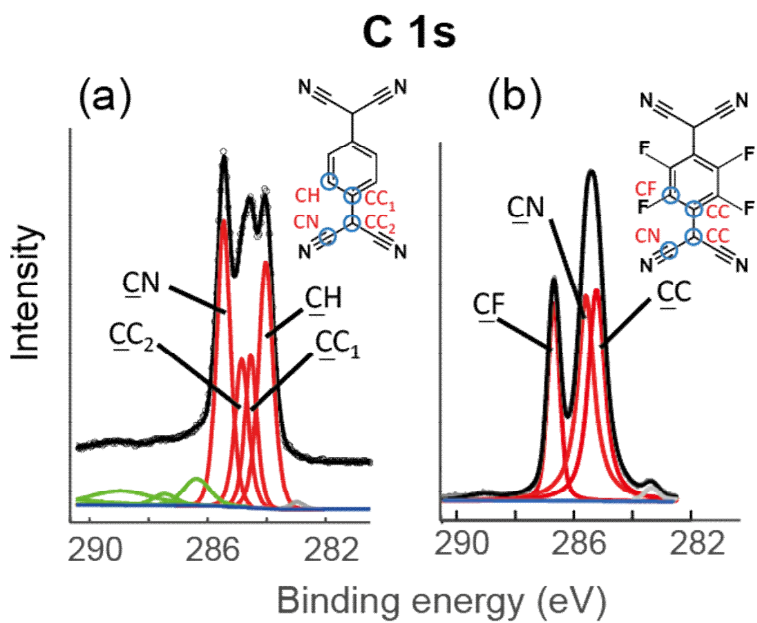

**Figure 2** SXPS results showing the C 1s spectra from the TCNQ:LDC phase (left) and the ordered phase of F$_4$TCNQ on Ag(100) (right), using a photon energy of 435 eV. Also shown are the different spectral components used to fit the experimental data (black). The main chemically shifted components are shown in red, the background in blue, C 1s shake-up satellites in green and a weak C 1s component associated with radiation damage in grey.

**NIXSW structure determination**

NIXSW experimental data were collected from the LDC and HDI1 phases of TCNQ on Ag(100), and from the ordered phase of F$_4$TCNQ on Ag(100), around the (200) Bragg condition. NIXSW relative photoemission intensity profiles were extracted for the N, F, and chemically distinct C components from these spectra, although the separate CC$_1$ and CC$_2$ components seen in the SXP spectrum from TCNQ could not be resolved at the higher photon energy of the NIXSW experiment. NIXSW photoemission-derived absorption profiles can be fitted (taking account of non-dipole effects in the angular dependence to the photoemission[16]) uniquely by two parameters, the coherent fraction, $f$, and the coherent position, $p$. In the idealised situation of an absorbing atom occupying a single well-defined site with no static or dynamic disorder ($f = 1.0$), the coherent position (expressed in units of the Bragg plane spacing, $d$) can be related to the height of this absorber site above the Bragg planes, $D = (p + n)d$, where $n$ is an integer (usually 0 or 1) chosen to ensure the implied interatomic distances are physically reasonable.[16] The coherent fraction is commonly regarded as an order parameter; including the effects of atomic vibrational amplitudes and some static disorder can lead to $f$ values as low as

~0.70,[25] but much lower values can only be attributed to the contributions of at least two distinctly different absorber heights. Table 1 shows the $f$ and $D$ values obtained from the NIXSW data from the two commensurate phases, TCNQ:LCD and $F_4$TCNQ; data for the incommensurate TCNQ:HDI1 phase are also included

**Table 1** Summary of the values of the structural parameters extracted from the NIXSW measurements from the commensurate phases of TCNQ:LCD and $F_4$TCNQ, together with the results of measurements from the TCNQ:HDI1 phase. Precision estimates in the final decimal place are shown in parentheses. The values for N obtained from the $F_4$TCNQ adsorption phase are discussed further in the text.

| component | TCNQ:LDC | | TCNQ:HDI1 | | $F_4$TCNQ | |
|---|---|---|---|---|---|---|
| | $f$ | $D$ (Å) | $f$ | $D$ (Å) | $f$ | $D$ (Å) |
| CH/CF | 0.68(10) | 2.70(5) | 0.67(10) | 2.69(5) | 0.79(10) | 3.01(5) |
| CC | 0.79(10) | 2.65(5) | 0.76(10) | 2.56(5) | 0.72(10) | 2.94(5) |
| CN | 0.70(10) | 2.51(5) | 0.69(10) | 2.45(5) | 0.56(10) | 2.71(5) |
| N | 0.81(10) | 2.36(5) | 0.63(10) | 2.28(5) | 0.20(10) | 2.90(20) |
| F | | | | | 0.56(10) | 3.04(5) |

For both of the TCNQ adsorption phases investigated by NIXSW, these data indicate adsorption of the molecule in the inverted bowl conformation, the N atoms being approximately 0.4 Å lower on the surface than the quinoid CH C atoms; the CC C atoms are slightly lower than these CH atoms and the CN C atoms slightly lower still. All the coherent fraction values are sufficiently high that it seems likely that all the atoms of each species have the same equilibrium height with no evidence that the mirror symmetry of the molecular conformation is significantly reduced by the adsorption. Any differences in the height of the molecule between the TCNQ:LDC and TCNQ:HDI1 phase are of marginal significance but seem to suggest the TCNQ molecule is more strongly bent in the incommensurate TCNQ:HDI1 phase.

The results for the adsorbed $F_4$TCNQ molecule, however, are significantly different. The high coherent fraction for the CF C atoms indicates that the quinoid ring is parallel to the surface, but some 0.3 Å higher than in the two TCNQ adsorption phases. However, the very low coherent fraction for the N atoms clearly indicates that the N atoms occupy at least two distinctly different heights; if there are just two such equally occupied heights, the reduction of the value of $f$ to only ~30% of the values for the CF and CC atoms would imply a height difference of ~0.8 Å [25]. Notice that with such a low value of $f$ the corresponding value of $D$, which may represent some weighted average of the contributing heights, is likely to have a significantly lower true precision than that implied by the statistical fitting routine.[25] This height difference would be expected to lead to a smaller height difference for the CN C atoms and a consequential smaller reduction in the associated $f$ value, consistent with the experimental result. The origin of the slightly reduced value of $f$ for the F atoms is less clear but may be due to the presence of a second F species, possibly atomic F, resulting from the radiation damage evidenced by the higher resolution SXP F 1s spectrum of Figure S3.

The main conclusions are that the central ring of $F_4TCNQ$ sits significantly higher above the Ag(100) surface than that of TCNQ, but while TCNQ adopts a symmetrical inverted bowl configuration, the cyano ends of the $F_4TCNQ$ are twisted such that some CN bonds point down towards the surface while others point up away from the surface. This latter conformation is similar to that adopted by TCNQ on Ag(111), the origin of the twisting in that case having been shown to be due to the presence of Ag adatoms in the molecular layer.[15] This clearly leads to the possibility that a similar adatom structure is associated with the Ag(100)-$F_4TCNQ$ surface. By contrast, the more symmetrical conformation of the TCNQ:LDC and TCNQ:HDI1 phases, together with their higher packing density, indicates that Ag adatom incorporation is probably not involved in these phases.

**Structure characterisation by Density Functional Theory**

To understand and determine the structure more completely, dispersion-inclusive DFT calculations were performed to determine the minimum energy configurations of the two commensurate phases investigated, namely Ag(100)-TCNQ LDC $\begin{pmatrix} 3 & 1 \\ -2 & 3 \end{pmatrix}$ and Ag(100)-$F_4TCNQ$ $\begin{pmatrix} 4 & 2 \\ -2 & 2 \end{pmatrix}$. DFT calculations were performed with the all-electron numeric atomic orbital code FHI-aims[26]. Dispersion interactions were modelled using both the Tkatchenko-Scheffler vdW$^{surf}$ method (PBE+vdW$^{surf}$)[27] and the non-local many-body dispersion method (PBE+MBD-NL) [28] implemented in the FHI-aims package (for more computational details see Methods section). For each of these adsorption phases, calculations were performed for two alternative models, one assuming there is no Ag adatom in the surface mesh and an alternative model that includes an Ag adatom. Table 2 shows a comparison of the NIXSW experimental parameter values with equivalent values extracted from the alternative model structure for the TCNQ:LDC surface based on PBE+MBD-NL DFT calculations. Note that calculations for possible starting structures for TCNQ coadsorbed with Ag adatoms failed to converge, indicating that no such stable structure exists. This is consistent with the qualitative evaluation of the STM, LEED, and NIXSW data in the previous section regarding the likely structure of the TCNQ:LDC phase and is further evidence that no adatoms are present in this structure.

**Table 2** Experimental NIXSW parameter values of the TCNQ:LDC phase on Ag(100) compared with values obtained from the PBE+MBD-NL DFT calculations for a structural model without Ag adatoms.

| Component | TCNQ:LDC Expt | | TCNQ:LDC DFT (no adatoms) | |
|---|---|---|---|---|
| | $f$ | $D$ (Å) | $f$ | $D$ (Å) |
| <u>C</u>H | 0.68(10) | 2.70(5) | 1.00 | 2.73 |
| <u>C</u>C | 0.79(10) | 2.65(5) | 0.98 | 2.65 |
| <u>C</u>N | 0.70(10) | 2.51(5) | 1.00 | 2.49 |
| N | 0.81(10) | 2.36(5) | 0.99 | 2.31 |

As shown in Table 2, the atomic layer spacings found for the minimum energy structure obtained from the DFT+MBD-NL calculations for the LDC TCNQ adsorption phase, assuming that no Ag adatoms are present, are in excellent agreement with the experimental values. A similar comparison with the results of DFT PBE+vdW$^{surf}$ calculations is shown in Table S2. Both dispersion correction methods provide adsorption geometries that are in close agreement with experiment. Note that $D$ values obtained in NIXSW are relative to the extended Bragg planes of the bulk structure. The theoretical values reported here are relative to the *average* height of the outermost Ag layer (which is slightly rumpled). This mode of comparison avoids problems associated with the slightly different lattice parameter of the bulk Ag crystal in the DFT calculations, and small multilayer spacing relaxations in the thin slab. The theoretical $f$ values in this table only take account of the reduction of this parameter (from unity) due to the small range of different heights of chemically equivalent but symmetrically inequivalent atoms in the optimised structure. These values of $f$, computed in this way, are inevitably up to ~30% higher than the experimental values because they take no account of static and dynamic disorder in the overlayer and the substrate [25]. Figure 3 shows the DFT-optimised structure of this Ag(100)-TCNQ:LDC $\begin{pmatrix} 3 & 1 \\ -2 & 3 \end{pmatrix}$ phase.

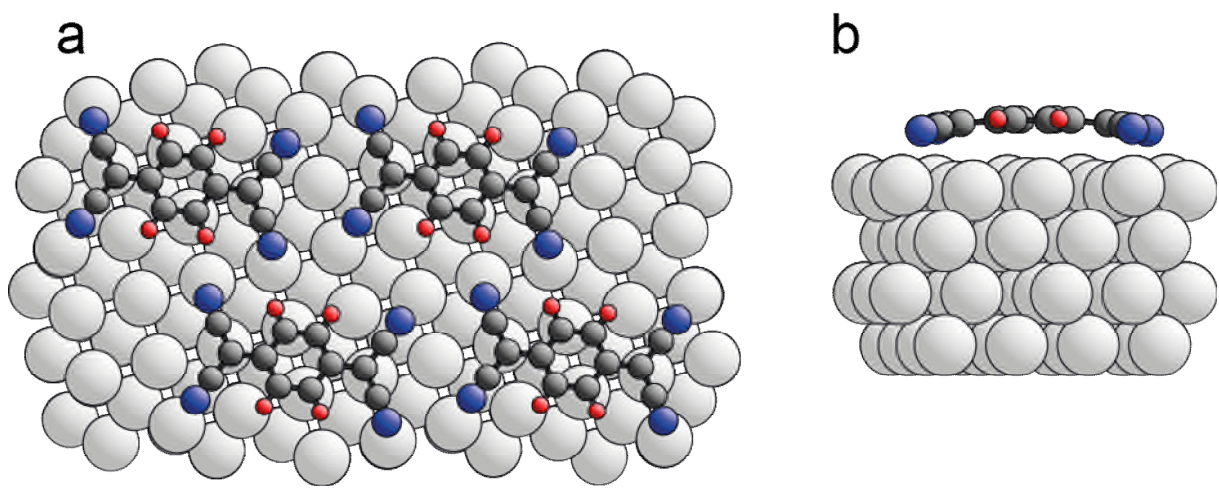

**Figure 3** (a) shows a top view of the DFT-optimised structure of the TCNQ:LDC phase on Ag(100). (b) shows a side view of a single adsorbed molecule within this adsorption phase. Adsorbate atom colouring: C, dark grey; H, red; N; blue.

Similar DFT calculations were also performed for two alternative models of the $F_4TCNQ$ $\begin{pmatrix} 4 & 2 \\ -2 & 2 \end{pmatrix}$ adsorption phase, in order to investigate the possible presence of Ag adatoms; one of these model structures contains only a single $F_4TCNQ$ molecule per surface unit mesh and the other contains one $F_4TCNQ$ and one Ag adatom per surface unit mesh. Table 3 shows a comparison of the experimental NIXSW parameter values with those predicted by DFT+MBD-NL calculations for the two optimised structural models. DFT+vdW$^{surf}$ calculations were also performed, the results being presented in Table S4 of the Supporting Information; these are in close agreement with MBD-NL results.

**Table 3** Comparison of experimental NIXSW parameter values of the $F_4TCNQ$ $\begin{pmatrix} 4 & 2 \\ -2 & 2 \end{pmatrix}$ phase on Ag(100) compared with values obtained from the DFT calculations for two alternative

structural models, with and without Ag adatoms. Theoretical values reported in the table are at the DFT+MBD-NL level.

| component | F$_4$TCNQ – Expt. | | F$_4$TCNQ – DFT no adatom | | F$_4$TCNQ – DFT with adatom | |
|---|---|---|---|---|---|---|
| | $f$ | $D$ (Å) | $f$ | $D$ (Å) | $f$ | $D$ (Å) |
| CF | 0.79(10) | 3.01(5) | 1.00 | 3.03 | 0.99 | 3.02 |
| CC | 0.72(10) | 2.94(5) | 0.94 | 2.88 | 0.98 | 2.93 |
| CN | 0.56(10) | 2.71(5) | 0.98 | 2.55 | 0.89 | 2.79 |
| N | 0.20(10) | 2.90(20) | 0.98 | 2.27 | 0.68 | 2.71 |
| F | 0.56(10) | 3.04(5) | 1.00 | 3.03 | 0.95 | 3.02 |

The agreement between experimental and computed $D$ values is clearly better for the with-adatom model than the no-adatom model. The most striking structural difference between the predictions for the two F$_4$TCNQ models is in the height range of the N atoms; without Ag adatoms, all the N atoms have almost identical heights (within a range of 0.04 Å), whereas with adatoms their height differs by up to 0.70 Å. This is attributable to the fact that while three N atoms in the molecule are adjacent to Ag adatoms (with N-Ag adatom distances in the range 2.33 Å to 2.40 Å and N atom heights above the underlying Ag(100) surface of 2.74-2.90 Å), the fourth N atom has no such adatom neighbour and bonds to an underlying Ag surface atom at a separation of 2.40 Å. This leads to the significant (31%) reduction in the theory value of $f$ in qualitative agreement with the experimental result. Quantitative agreement with this value is poor but estimating the true precision in both $f$ and $D$ when the nominal value of $f$ is as low as 0.20 is very difficult, as we have demonstrated elsewhere.[25] This behaviour of the N atom heights strongly favours the adatom model shown in Figure 4; the different N atom heights lead to a twisting of the cyano endgroups, but the central quinoid ring remains parallel to the surface. The Ag adatoms occupy 4-fold coordinated hollow sites on the surface. Predicted $D$ values for most of the other atoms show generally comparable agreement with experiment for both models; for the CN C atoms the agreement is clearly superior for the with-adatom model, although neither model predicts any significant lowering of the associated $f$ value.

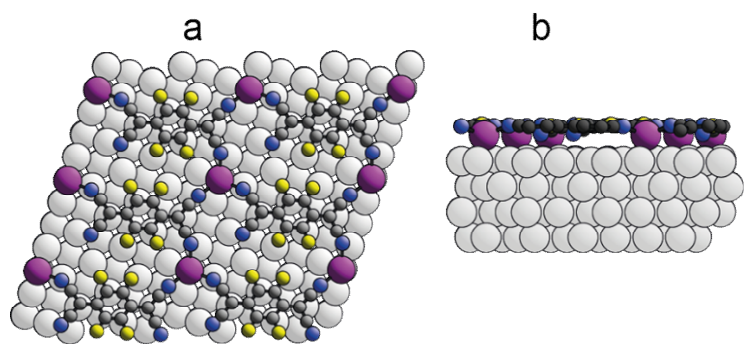

**Figure 4** Optimised structure of the adatom model of Ag(100)-F$_4$TCNQ in (a) top view and (b) side view. The Ag adatoms are shown in purple. Adsorbate atom colouring: C, dark grey; F, yellow; N; blue.

In summary, a comparison of the structural data from the NIXSW experiments and the DFT calculations shows clearly that the TCNQ:LDC phase on Ag(100) does not involve Ag adatoms whereas the similarly commensurate ordered phase of F$_4$TCNQ on Ag(100) does involve Ag adatoms. Furthermore, TCNQ on Ag(100) in the LDC phase adopts an inverted bowl configuration, whereas F$_4$TCNQ on Ag(100) adopts a twisted conformation in which three of its CN groups bond to Ag adatoms while the remaining CN group bonds to the Ag surface. The comparison of the NIXSW results from the TCNQ:LDC and TCNQ:HDI1 phase also indicates that TCNQ adopts the inverted bowl configuration in both of these phases. Thus, TCNQ at room temperature does not induce the extraction of substrate atoms, but its fluorinated counterpart, F$_4$TCNQ, does. However, annealing of the TCNQ:LDC phase leads to the formation of the TCNQ:W2 phase, the STM images of which are tentatively interpreted as indicative of the presence of Ag adatoms. In order to rationalise these observations, we now explore the thermodynamic driving forces for adatom network formation and the implications of these differences for the surface electronic structure.

**Adsorption energies with and without Ag adatoms**

Decomposition of the energetics of the molecular adsorption offers a possible way to understand why the adatom network might be more favourable in some cases, but not in others. To calculate the adsorption energy, $E_{ads}$, for TCNQ and F$_4$TCNQ on Ag(100) in the absence of Ag adatoms we used Equation 1, where $E_{TS}$ is the energy of the total system, $E_{surf}$ is the energy of the bare, clean surface and $E_{mol}$ is the energy of the isolated molecule. However, when adatoms are present, the number of Ag atoms in the complete substrate+molecule+adatom slab is increased by one; so, in order to calculate the adsorption energy, one must also subtract from $E_{TS}$ the energy of a single isolated atom, $E_{Ag}$, and the energy cost of removing this atom from the substrate (which is equal to the cohesive energy of crystalline Ag, $E_{coh}$). The final result is given by Equation 2, where we note that, by convention, $E_{coh}$ and $E_{ads}$ are defined as positive, whereas all the other energy terms are negative.

$$E_{ads} = E_{TS} - (E_{surf} + E_{mol}) \tag{1}$$

$$E_{ads} = E_{TS} - (E_{surf} + E_{mol} + E_{Ag} - E_{coh}) \tag{2}$$

Adsorption energies per unit surface area are summarised in Table 4, calculated using both the DFT+MBD-NL and DFT+vdW$^{surf}$ levels. Energies computed at the DFT+MBD-NL level are lower than those calculated at the DFT+vdW$^{surf}$ level in all cases. This is known to be a general feature of these methods and is to be expected, as many-body dispersion corrects for some of the adsorption energy over-binding of the pairwise additive DFT+vdW$^{surf}$ scheme[29]. As we already noted, both methods provide very similar predictions for the adsorption heights, but DFT+MBD-NL is expected to provide a more accurate description of adsorption energies. The results presented in Table 4 show that for F$_4$TCNQ on Ag(100) adatom incorporation is favoured, consistent with our structural conclusions. Taking the more reliable energy values of the DFT+MBD-NL calculations, we see that the energetic advantage of the adatom structure for F$_4$TCNQ adsorption is 0.67 meV/nm$^2$.

**Table 4** Adsorption energies (eV/nm$^2$) of Ag(100)-TCNQ in the LDC phase and Ag(100)-F$_4$TCNQ with and without adatoms.

| Adsorbate | no adatoms DFT+MBD-NL | no adatoms DFT+vdW$^{surf}$ | with adatoms DFT+MBD-NL | with adatoms DFT+vdW$^{surf}$ |
|---|---|---|---|---|
| TCNQ | 4.04 | 4.96 | N/A | N/A |
| F$_4$TCNQ | 3.86 | 5.63 | 4.52 | 5.78 |

As shown in Table S1, the TCNQ LDC phase has a significantly higher molecular packing density than the F$_4$TCNQ $\begin{pmatrix} 4 & 2 \\ -2 & 2 \end{pmatrix}$ phase, so it is interesting to consider the energetics in the hypothetical situation in which TCNQ adopts the larger $\begin{pmatrix} 4 & 2 \\ -2 & 2 \end{pmatrix}$ unit mesh. The results of these calculations are shown in Table S4 of the Supporting Information. There are two significant conclusions: first, the adsorption energy at the MBD-NL level of TCNQ in this larger mesh (in the absence of Ag adatoms) is lower by 0.95 eV/nm$^2$ than in the LDC mesh, and second the adsorption energy is even lower (but only by 0.18 eV/nm$^2$) if Ag adatoms are incorporated into the structure. The conclusion is clearly that TCNQ prefers the small LDC mesh and even if it adopted the larger mesh to allow more space for Ag adatom incorporation, this would still not be energetically favoured.

Some insight into the reasons for the energetic differences shown in Table 4 may be gained by separating the individual contributions to these total adsorption energies, namely the energy cost of the deformation of the Ag(100) surface and of the gas-phase molecule, E$_{deformation}$, and the energy gain resulting from the bonding interactions, E$_{interaction}$. The results of these calculations are presented in Table 5, computed at the PBE+MBD-NL level. PBE+vdW$^{surf}$ calculations show the same qualitative trends. Perhaps unsurprisingly, the energy cost of the deformation is lowest for the TCNQ LDC phase in which there is no metal surface reconstruction, and the molecule simply shows the weak bending into the inverted bowl conformation. The deformation energy is higher for both models of the Ag(100)-F$_4$TCNQ system, but is by far the highest for the model of this structure that includes Ag adatoms, although this energy cost is more than offset by a large increase in the interaction energy.

**Table 5** Adsorption energy decomposition (eV/nm$^2$) of Ag(100)-TCNQ and Ag(100)-F$_4$TCNQ with and without adatoms computed at the PBE+MBD-NL level.

|  | Ag(100)-TCNQ | Ag(100)-F$_4$TCNQ | |
| --- | --- | --- | --- |
|  | no adatom | no adatom | with adatom |
| $E_{ads}$ | 4.04 | 3.86 | 4.52 |
| $E_{deformation}$ | −0.23 | −1.11 | −1.60 |
| $E_{interaction}$ | 4.27 | 4.97 | 6.12 |

Further insight into the nature of the F$_4$TCNQ plus adatom phase is provided by an alternative breakdown of the total adsorption energy to determine what fraction of the total adsorption energy is associated with adsorbate-substrate interaction and what fraction is associated with lateral intralayer interaction. The results (computed at the PBE+MBD-NL level) are presented in Table 6, which shows clearly that the intralayer interaction contributes very significantly (almost 30%) to the total adsorption energy for F$_4$TCNQ adsorption. F$_4$TCNQ interacts strongly with the Ag adatoms to produce a two-dimensional metal-organic framework with a strong lateral cohesion. This appears to be the key driving force for the reconstruction. We have recently observed a similar 2D MOF formation in the case of TCNQ coadsorbed with potassium ions on Ag(111).[30]

**Table 6** Energetic breakdown of the total adsorption energy into interactions between the adsorbate and the substrate and interactions within the layer computed at the PBE+MBD-NL level.

|  | Ag(100)-F$_4$TCNQ |
| --- | --- |
|  | with adatoms |
| $E_{ads}$ (eV/nm$^2$) | 4.52 |
| Adsorbate – substrate interaction (eV/nm$^2$) | 3.19 |
| Adsorbate – substrate interaction (%) | 71% |
| Intralayer interaction (eV/nm$^2$) | 1.33 |
| Intralayer interaction (%) | 29% |

While it is clear that, for TCNQ adsorption at room temperature, the resulting commensurate LDC phase does not include Ag adatoms, it is interesting to consider the evidence from the STM images that the TCNQ:W2 phase may include Ag adatoms. If this were to be the case it may be that the TCNQ:LDC phase, without adatoms, is effectively a metastable phase, and that annealing and the associated increase in atomic mobility may lead to a true equilibrium phase that does include Ag adatoms. Because the TCNQ:W2 phase is incommensurate, it is not possible to perform a DFT calculation of this exact structure, but calculations using a model commensurate structure of similar size may provide some insight into the likely behaviour of the TCNQ:W2 phase. The results of DFT calculations of this type are reported in the Supporting Information (see Figure S5, Table S5, and accompanying text) but prove to be inconclusive. In particular, they show that the structure that is energetically favoured does incorporate Ag adatoms, but the energy difference is very marginal and, in particular, is less than *kT* at room temperature. The implication that structures with and without Ag adatoms should coexist could be consistent with the fact that STM images produced after annealing do show coexistence of the TCNQ:W2 and TCNQ:W1 structures, with the TCNQ:W2 apparently containing Ag adatoms whereas the TCNQ:W1 phase appears to have less, or no, Ag adatoms.

**Surface electrostatics**

Prototypical electron acceptors such as TCNQ and F$_4$TCNQ lead to significant charge rearrangement and dipole formation upon surface adsorption, and adatom formation can have a very significant impact. An increase in the work function relative to the clean surface is found for all calculated systems, as expected for adsorption of electron acceptor molecules. Changes to the system work function can be analysed in terms of contributions arising from electrostatic dipole effects and chemical bonding , as performed by Hofmann *et al.* [31]. As shown in equation 3, we can decompose the change in work function ΔΦ (with respect to the clean surface) into two separate contributions. One contribution arises from the drop in electrostatic potential due to the molecular dipole in the adsorbed molecular geometry, $\Delta E_{\text{mol}}$. The second term arises from charge rearrangements due to the electronic interaction of the molecule and substrate, $\Delta E_{\text{bond}}$.

$$\Delta \Phi = \Delta E_{\text{mol}} + \Delta E_{\text{bond}} \qquad (3)$$

Table 7 shows the values of the work function increase in all three adsorption phases, with the largest increase caused by Ag(100)- F$_4$TCNQ without adatoms. In all three cases, this increase is due to a strong electrostatic potential contribution due to the dipole perpendicular to the surface of the distorted molecule . This is only partially compensated by the reduction of the dipole due to the electronic interaction of adsorbate and substrate, which is a measure of the strength of local interactions between adsorbate and substrate. *ΔE*$_{\text{bond}}$ is much larger for F$_4$TCNQ than for TCNQ. In the bent adsorbate geometry of F$_4$TCNQ in the absence of Ag adatoms, PBE+MBD-NL calculations predict a molecular dipole perpendicular to the surface of 1.98 Db, whereas including the Ag adatom leads to a much smaller dipole moment of 0.89 Db. The molecular geometry in the presence of adatoms is less distorted (see Figure 4), leading to smaller associated *ΔE*$_{\text{mol}}$ of –0.24 eV. Simultaneously, the interface dipole of opposite sign due to bond formation is also reduced. The net work function change is 0.21 eV lower than in the no-adatom case. Therefore, the formation of an adatom-incorporated F$_4$TCNQ layer reduces the increase in work function, which would otherwise result from adsorption of the strong electron acceptor F$_4$TCNQ, to a value that is closer to that of TCNQ on Ag(100).

**Table 7** Work function changes computed for both molecular adsorbates relative to the computed value for the clean surface of 4.19 eV. Shown are the total change in work function,

$\Delta\Phi$, the electrostatic contribution of the work function, $\Delta E_{mol}$, and the contribution to the work function due to chemical interaction $\Delta E_{bond}$. All values calculated at the PBE+MBD-NL level.

| Work function change and components | Ag(100)-TCNQ:LDC | Ag(100)-F$_4$TCNQ no adatom | Ag(100)- F$_4$TCNQ with adatom |
|---|---|---|---|
| $\Delta\Phi$ (eV) | 0.62 | 0.80 | 0.59 |
| $\Delta E_{mol}$ (eV) | −0.45 | −0.59 | −0.24 |
| $\Delta E_{bond}$ (eV) | 1.07 | 1.39 | 0.83 |

**CONCLUSIONS**

The interface between metal substrates and organic adsorbates plays a key role in determining the functional properties of organic electronic devices. Since strong electron acceptor molecules play a significant role in organic electron materials, it is important to gain an understanding of the various interactions, at the interface, which result from deposition of these molecules. Previous work has shown that both TCNQ and F$_4$TCNQ adsorption on noble metal surfaces can, in some systems, lead to a surface reconstruction, the resulting overlayer being a 2D MOF containing the adsorbed molecule and adatoms of the underlying metal species. However, in other cases it appears that no such reconstruction occurs. Through joint experimental and theoretical techniques, we have shown that the commensurate adsorption phases of these two molecules on Ag(100) lie on either side of the condition for this reconstruction. Specifically, F$_4$TCNQ does lead to the formation of an adatom-incorporated overlayer, whereas TCNQ does not. Moreover, we present the first clear quantitative experimental structural evidence that TCNQ adopts the inverted bowl conformation on the surface that had previously been predicted by DFT calculations to occur for both of these molecular electron acceptors on a range of surfaces.

Our analysis of the underlying energetics provides valuable insight into the factors that determine whether adsorption of a molecular electron acceptor does, or does not, cause incorporation of metal adatoms into the overlayer. Specifically, our calculated adsorption energies clearly favour Ag adatom incorporation when F$_4$TCNQ is adsorbed. Furthermore, comparison of DFT and experimental NIXSW structural parameter values are fully consistent with the conclusions that TCNQ adsorption at room temperature does not lead to Ag adatom incorporation into the overlayer, whereas adsorption of $F_4TCNQ$ adsorption under similar conditions does lead to Ag adatom incorporation. A breakdown of the contributions to these adsorption energies shows that it is the strong interaction of $F_4TCNQ$ with Ag adatoms resulting in the formation of a true 2D MOF that determines the reconstruction behaviour.

Despite the clear evidence, both experimentally and computationally, that TCNQ does not create a commensurate adatom-incorporated overlayer when adsorbed at room temperature, annealing to elevated temperatures does produce an incommensurate phase that appears, on the basis of qualitative STM image evaluation, to involve incorporation of Ag adatoms. We infer that in this case kinetic barriers preclude the formation of this phase at room temperature. However, the fact that this phase cannot be formed without coexistence of areas of other phases that appear not to incorporate Ag adatoms suggests that the energetics determining whether adatom incorporation occurs may be marginal.

Finally, we show that the incorporation of Ag adatoms into the $F_4TCNQ$ overlayer significantly modifies the interface electronic properties. Stronger acceptors are expected to lead to larger work function increases. However, the stronger acceptor $F_4TCNQ$ also shows a stronger driving force for 2D MOF formation with Ag adatoms, which counteracts work function increase. Therefore, our results suggest that the maximal work function increase that can be achieved by acceptor adsorption on a Ag(100) surface may be naturally limited by the formation of Ag adatom networks.

**Supporting Information**: Additional experimental results: LEED, STM, SXPS and NIXSW; Additional DFT structural and energetics results: TCNQ and $F_4TCNQ$ on Ag(100).

**ACKNOWLEDGEMENTS**
The authors thank Diamond Light Source for allocations SI17261 and SI20785 of beam time at beamline I09 that contributed to the results presented here. P.T.P.R. and P.J.B. acknowledge financial support from Diamond Light Source and EPSRC. *G.C.* acknowledges financial support from the EU through the ERC Grant "VISUAL-MS" (Project ID: 308115). B.S. and R.J.M. acknowledge doctoral studentship funding from the EPSRC and the National Productivity Investment Fund (NPIF). R.J.M. acknowledges financial support *via* a UKRI Future Leaders Fellowship (MR/S016023/1). We acknowledge computing resources provided by the EPSRC-funded HPC Midlands+ Computing Centre (EP/P020232/1) and the EPSRC-funded Materials Chemistry Consortium (EP/R029431/1) for the ARCHER2 U.K. National Supercomputing Service (http://www.archer2.ac.uk).

TOC

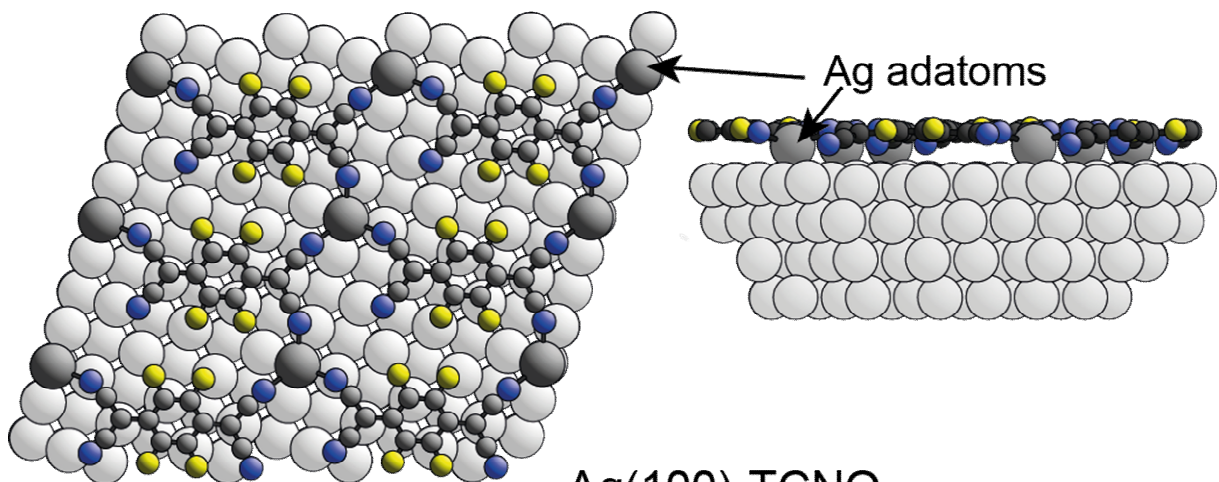
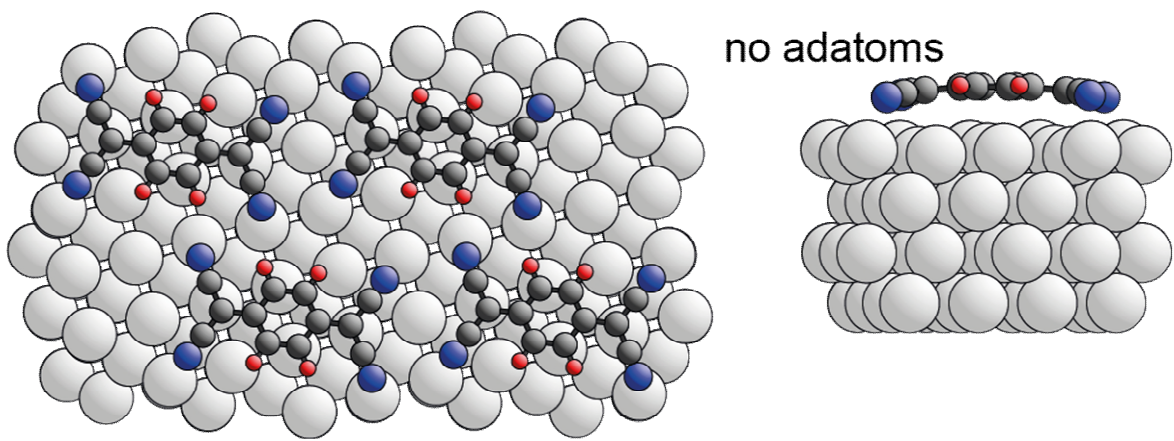

# Supporting Information to

# Thermodynamic Driving Forces for Substrate Atom Extraction by Adsorption of Strong Electron Acceptor Molecules

**Paul Ryan[1,2], Philip James Blowey[1,3], Billal S. Sohail[4], Luke A. Rochford[1], David A. Duncan[1], Tien-Lin Lee[1], Peter Starrs [1,5], Giovanni Costantini[4], Reinhard J. Maurer[4]\***
**, David Phillip Woodruff[3]**Error! Bookmark not defined.

*(1) Diamond Light Source, Harwell Science and Innovation Campus, Didcot, OX11 0DE, UK*
*(2) Department of Materials, Imperial College, London SW7 2AZ, UK*
*(3) Department of Physics, University of Warwick, Coventry CV4 7AL, UK*
*(4) Department of Chemistry, University of Warwick, Coventry CV4 7AL, UK*
*(5) School of Chemistry, University of St. Andrews, St. Andrews, KY16 9AJ, UK*

**Contents**

1. Additional experimental results: LEED, STM, SXPS and NIXSW
2. Additional DFT structural and energetics results: TCNQ and F$_4$TCNQ on Ag(100)

### 1. Additional experimental results: LEED, SXPS and NIXSW

STM images and LEED patterns recorded from all of the ordered phases of TCNQ and F$_4$-TCNQ on Ag(100) are shown in Figure S1, together with simulations using the LEEDpat program[1] based on the matrices reported in Table S1. The LEED patterns observed from surfaces containing the two coexisting windmill structures (TCNQ:W1 and TCNQ:W2) were complex, as may be expected from the coexistence of two such large unit mesh structures, so no attempt has been made to identify all the observed diffracted beams, but an example of the diffraction pattern is shown in Figure S1. Notice that an unusual feature of the LEED pattern from the F$_4$TCNQ overlayer is the splitting into groups of four spots of all beams with indices comprising one integer and one half-integer, such as (0 ½), (-1 -½) (but not (½ ½)). These beams are circled in Figure S1. The splitting is not reproduced in the simulated pattern using the commensurate matrix $\begin{pmatrix} 2 & -2 \\ 4 & 2 \end{pmatrix}$ and is attributable to antiphase domain boundaries on the overlayer structure. This assignment is supported by the observation that the splitting was found to gradually disappear with extended annealing.

---

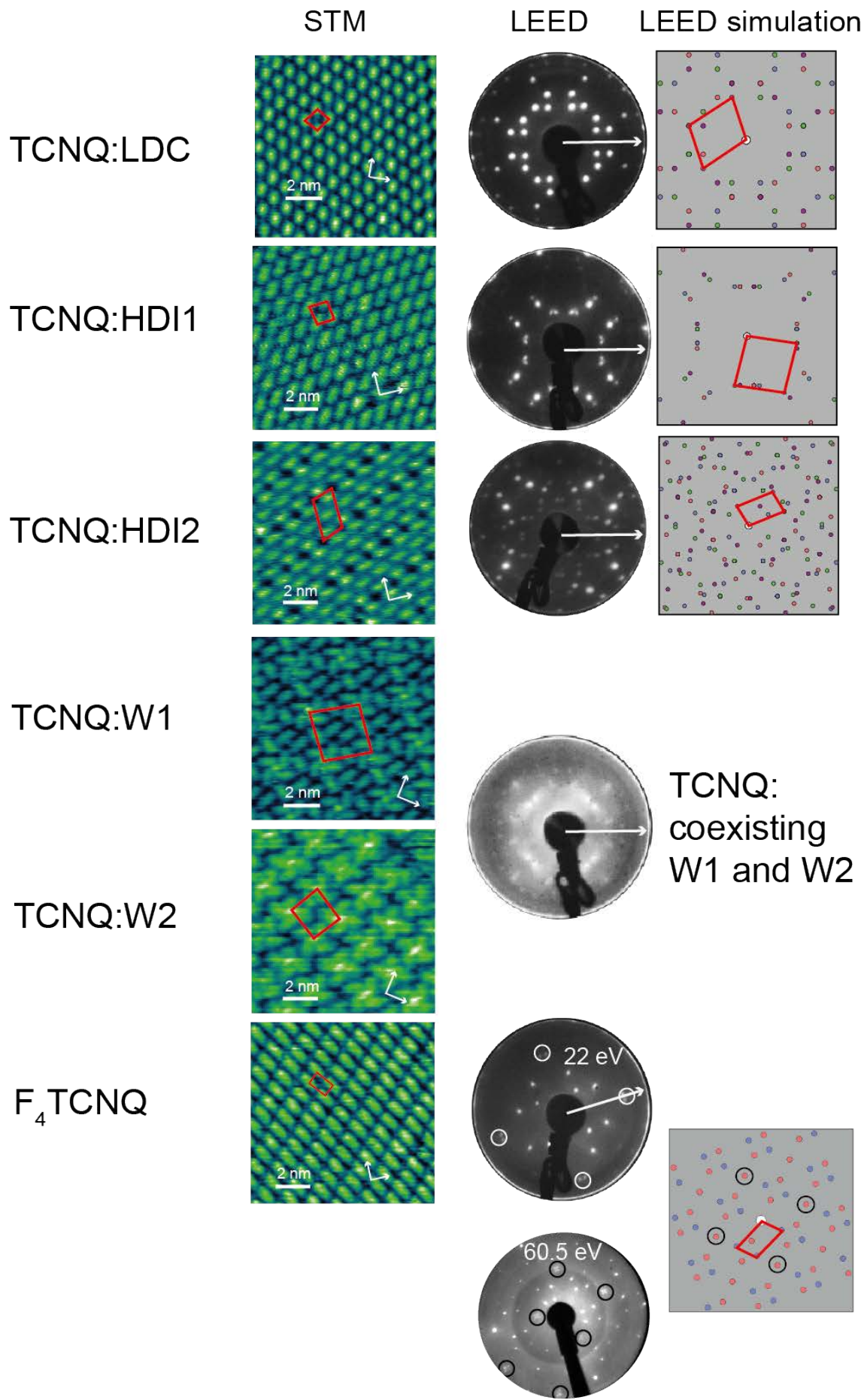

**Figure S1**. LEED patterns and simulated LEED patterns based on the matrices of Table S1, for the LDC, HDI1, HDI2 and, the coexisting W1 and W2 phases of TCNQ on Ag(100) and the ordered phase of F$_4$TCNQ on Ag(100). Electron energies are TCNQ:LDC, 18.0 eV; TCNQ:HDI1, 23.5 eV; TCNQ:HDI2, 16.0 eV; TCNQ:W1 & TCNQ:W2, 18.0 eV; F$_4$TCNQ, 22.0 eV and 60.5 eV. In the simulated LEED patterns diffracted beams arising from different rotational and mirror domains are shown in different colours. Unit meshes and corresponding reciprocal unit meshes are superimposed in red. <110> directions in the surface are shown as white arrows superimposed on the STM images and LEED patterns. STM tunnelling conditions (sample bias and tunnelling current): TCNQ:LDC, 0.25 V, 150 pA; TCNQ:HDI1, -0.10 V, 100 pA; TCNQ:HDI2, 1.00 V, 75 pA; TCNQ:W1, -0.05 V, 150 pA; TCNQ:W2, 1.25 V, 75 pA. F$_4$TCNQ, -1.00 V, 75 pA. The different conditions were chosen chosen to optimise imaging contrast.

**Table S1** Summary of the different ordered adsorption phases of TCNQ and F$_4$TCNQ found on Ag(100).

| Phase Descriptor | Matrix | Unit mesh area (Å$^2$) | No of molecules per unit mesh | Area per molecule (Å$^2$) | Preparation |
|---|---|---|---|---|---|
| TCNQ: LDC | $\begin{pmatrix} 3 & 1 \\ -2 & 3 \end{pmatrix}$ | 92 | 1 | 92 | Low coverage deposition |
| TCNQ: HDI1 | $\begin{pmatrix} 3.06 & 0.76 \\ 0.34 & -3.05 \end{pmatrix}$ | 80 | 1 | 80 | Higher coverage deposition |
| TCNQ: HDI2 | $\begin{pmatrix} 6.51 & 2.63 \\ -2.08 & 3.60 \end{pmatrix}$ | 240 | 2/3 | 120/80 | Anneal phase HDI1 to 280-340°C |
| TCNQ: W1 | $\begin{pmatrix} 7.84 & 5.09 \\ -6.01 & 6.22 \end{pmatrix}$ | 660 | 6 | 110 | Anneal phase LDC to 260-340°C (mixed phases) |
| TCNQ:W2 | $\begin{pmatrix} 2.9 & 5.7 \\ -6.1 & 2.6 \end{pmatrix}$ | 351 | 4 | 88 | Anneal phase LDC to 340°C (mixed phases) |
| F$_4$TCNQ | $\begin{pmatrix} 4 & 2 \\ -2 & 2 \end{pmatrix}$ | 100 | 1 | 100 | RT deposition |

Table S1 summarises the main parameters of these different ordered adsorption phases. Note that the numbers of molecules per unit mesh, and hence the area per molecule, are based on the assumption that each elongated bright feature in the STM images corresponds to a molecule. For the TCNQ:HDI2 phase, this value is somewhat ambiguous; there appear to be three bright features per unit mesh but attributing all of these to TCNQ molecules seems to lead to some unreasonably short intermolecular distances. The STM image also appears to show two rather different types of bright features, one of which shows a brighter circular centre that could, perhaps, be due to the presence of an Ag adatom rather than a TCNQ molecule.

Figure S2 shows an STM image recorded from the Ag(100)-TCNQ surface under conditions leading to the formation of islands of the W2 phase, such as that outlined in purple. Notice that the image also shows isolated and linear groups TCNQ 'windmill' structures with bright centres, possibly attributable to Ag adatoms.

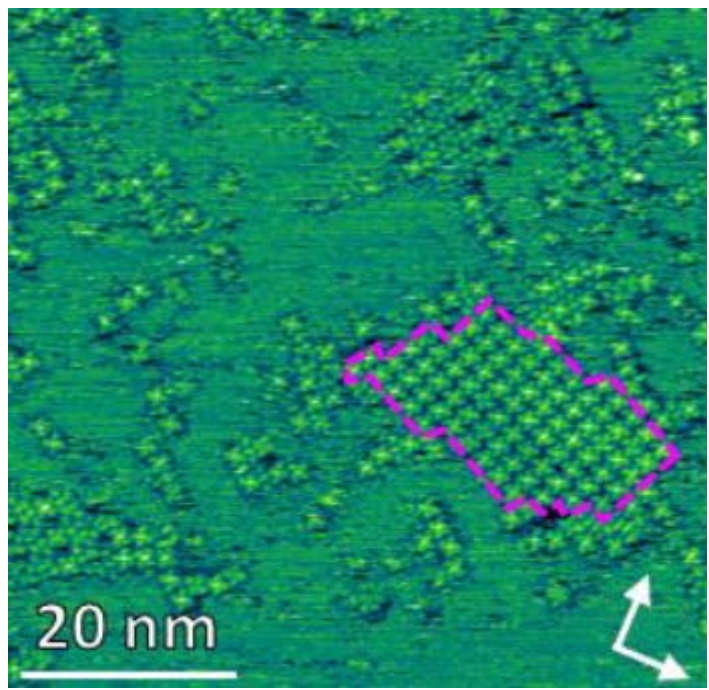

**Figure S2**. STM image of the Ag(100)-TCNQ surface prepared under conditions leading to the formation of islands of the W2 phase. One such island is outlined in purple. $V_{sample}$=1.25 V, tunnelling current 75 pA.

N 1s and F 1s SXP spectra from the TCNQ:LDC phase on Ag(100), and from the ordered phase of F$_4$TCNQ on Ag(100), which complement the C 1s spectra shown in Figure 2 of the main manuscript, are shown in Figure S3.

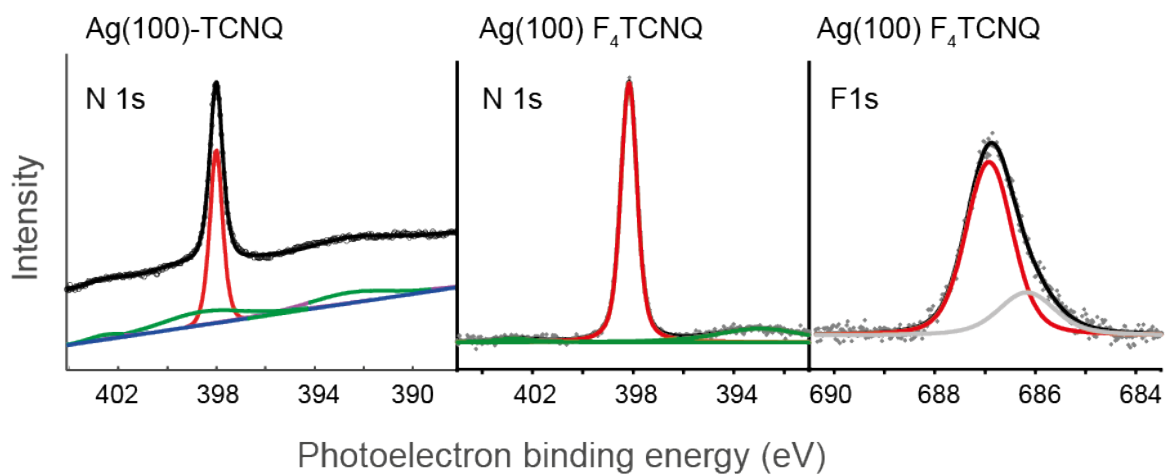

**Figure S3**. N 1s and F 1s SXP spectra from the LDC phase of TCNQ on Ag(100) and from the ordered phase of F$_4$TCNQ on Ag(100). The fits to the main peaks are shown in red, while satellites from the Ag 3d$_{3/2}$ peak are shown in green. An additional F 1s component attributed to a surface species arising from radiation damage is shown in grey.

From both surfaces the N 1s spectrum is dominated by a single peak, indicating that all N atoms are in chemically similar bonding sites. The F 1s spectrum does show a second component that is attributed to the effects of radiation damage. F is known to have a particularly high cross-section for electron and photon stimulated desorption, but the presence of this component implies that some F may remain on the surface, possibly in atomic form, rather than being removed into the vacuum.

Chemical-state specific NIXSW measurements of the TCNQ:LDC and TCNQ:HDI1 phases and the ordered phase of $F_4TCNQ$ were recorded using the C 1s, N 1s and F 1s photoemission signals as the photon energy was scanned through the normal incidence Ag(200) Bragg reflection condition, as described in the main manuscript. The experimental results and best theoretical fits are shown in Figure S4. The values of the coherent fractions and positions used in these best-fit theory curves for the TCNQ:LDC and $F_4TCNQ$ phases (and also from the TCNQ:HCI1 phase) are reported in Table 1 of the main paper.

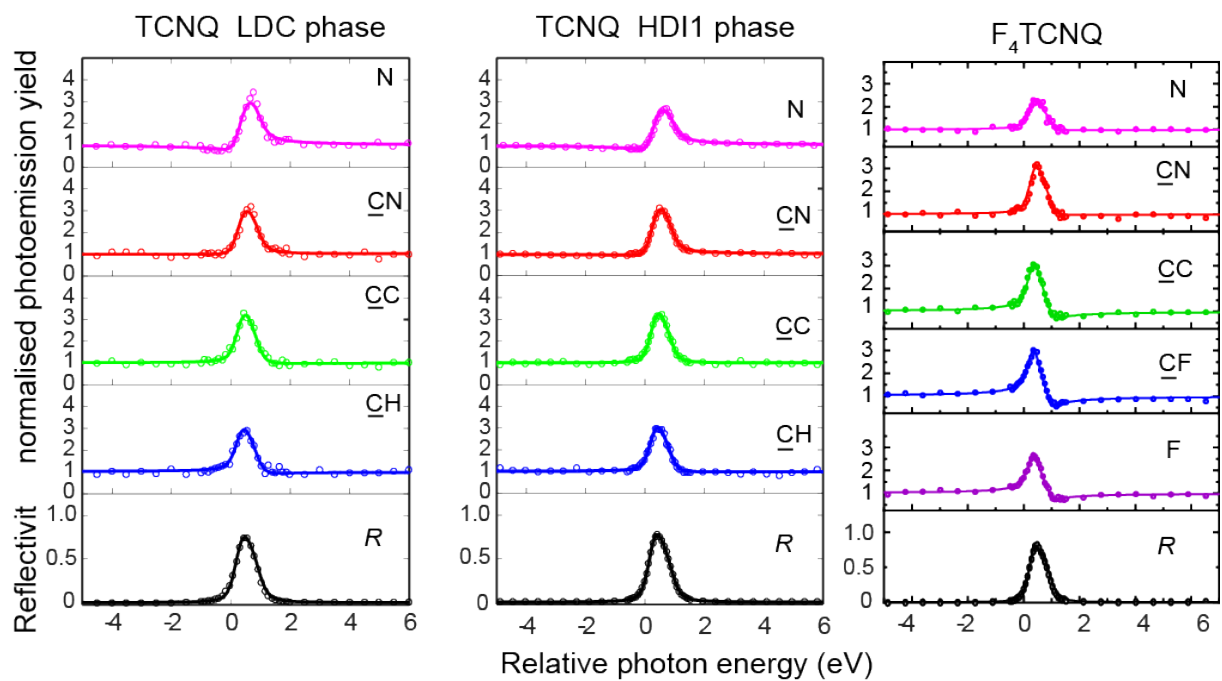

**Figure S4**. NIXSW photoemission yield curves obtained from samples corresponding to the TCNQ:LDC and TCNQ:HDI1 phases and the ordered phase of $F_4TCNQ$ on Ag(100) using the (200) reflection of the substrate. Photon energies are quoted relative to the Bragg energy of approximately 3036 eV. Least square fits (solid lines) to the photoemission yields (circles) were obtained to extract the values of the coherent fractions and coherent positions reported in Table 1 of the main manuscript. Also shown is the reflectivity, $R$.

## 2 Additional DFT structural and energetics results: TCNQ and F$_4$TCNQ on Ag(100)

The results of the DFT structural optimisation for the TCNQ:LDC phase on Ag(100) at the PBE+vdW$^{surf}$ level are shown below in Table S2 together with the experimental values. The results obtained from DFT+MBD-NL calculations of this phase are shown in Table 2 of the main paper. The two computational methods yield very similar results.

**Table S2** Comparison of experimental NIXSW parameter values of the TCNQ LDC phase on Ag(100) compared with values obtained from PBE+vdW$^{surf}$ calculations for a structural model that does not include Ag adatoms.

|           | TCNQ LDC Expt | | TCNQ LDC DFT (no adatoms) | |
|-----------|---------|---------|------|--------|
| Component | $f$     | $D$ (Å) | $f$  | $D$ (Å) |
| CH        | 0.68(10) | 2.70(5) | 1.00 | 2.76 |
| CC        | 0.79(10) | 2.65(5) | 0.98 | 2.66 |
| CN        | 0.70(10) | 2.51(5) | 1.00 | 2.50 |
| N         | 0.81(10) | 2.36(5) | 1.00 | 2.34 |

The results of the structural optimisation for F$_4$TCNQ on Ag(100), with and without Ag adatoms, at the DFT+MBD-NL level are reported in Table 3 of the main paper, while those obtained from similar DFT+vdW$^{surf}$ calculations are shown in Table S3.

**Table S3** Comparison of experimental NIXSW parameter values of the F$_4$TCNQ phase on Ag(100) compared with values obtained from DFT+vdW$^{surf}$ calculations for two alternative structural models, with and without Ag adatoms.

|           | F$_4$TCNQ – Expt. | | F$_4$TCNQ – DFT no adatom | | F$_4$TCNQ – DFT with adatom | |
|-----------|---------|---------|------|--------|------|--------|
| component | $f$     | $D$ (Å) | $f$  | $D$ (Å) | $f$  | $D$ (Å) |
| CF        | 0.79(10) | 3.01(5) | 1.00 | 2.96 | 0.99 | 2.92 |
| CC        | 0.72(10) | 2.94(5) | 0.98 | 2.85 | 0.98 | 2.85 |
| CN        | 0.56(10) | 2.71(5) | 1.00 | 2.56 | 0.96 | 2.77 |
| N         | 0.20(10) | 2.90(20) | 0.99 | 2.28 | 0.69 | 2.74 |
| F         | 0.56(10) | 3.04(5) | 0.97 | 2.94 | 0.97 | 2.95 |

As discussed in the main manuscript, one possible rationale for the fact that Ag adatom incorporation into the LDC phase of adsorbed TCNQ does not occur is that in this phase the high molecular packing density does not allow sufficient space. It is therefore interesting to know if Ag adatom would be favoured if adsorbed TCNQ were to adopt the larger $\begin{pmatrix} 4 & 2 \\ -2 & 2 \end{pmatrix}$ unit mesh adopted by $F_4$TCNQ. The results of these calculations are shown in Table S4.

**Table S4** Calculated value of the adsorption energy (eV/nm2) of TCNQ, were it to be adsorbed on Ag(100) in the larger $\begin{pmatrix} 4 & 2 \\ -2 & 2 \end{pmatrix}$ unit mesh adopted by $F_4$TCNQ on this surface, with and without adatoms computed at the PBE+MBD-NL and PBE+vdW$^{surf}$ level. For comparison the values obtained for adsorption in the LDC $\begin{pmatrix} 3 & 1 \\ -2 & 3 \end{pmatrix}$ unit mesh (Table 4 of the main manuscript) are included.

| Unit mesh | no adatoms DFT+MBD-NL | no adatoms DFT+vdW$^{surf}$ | with adatoms DFT+MBD-NL | with adatoms DFT+vdW$^{surf}$ |
|---|---|---|---|---|
| $\begin{pmatrix} 4 & 2 \\ -2 & 2 \end{pmatrix}$ | 3.09 | 5.41 | 2.91 | 5.44 |
| LDC $\begin{pmatrix} 3 & 1 \\ -2 & 3 \end{pmatrix}$ | 4.04 | 4.96 | N/A | N/A |

As remarked in the main manuscript, the STM images of the TCNQ:W2 phase do indicate that Ag adatoms may be incorporated in this phase, so DFT calculations were performed to try to cast light on this possibility. Because the TCNQ:W2 phase is incommensurate, it is not possible to perform a DFT calculation of this exact structure, but calculations using a model commensurate structure of similar size may provide some insight into the likely behaviour of the TCNQ:W2 phase. STM images of a commensurate $\begin{pmatrix} 6 & 3 \\ -3 & 6 \end{pmatrix}$ phase formed by coadsorption of TCNQ with Cs [2] are very similar to those of the W2 phase, so we choose this unit mesh for our calculation. We therefore modelled a set of TCNQ adsorption structures incorporating these molecular windmills using this $\begin{pmatrix} 6 & 3 \\ -3 & 6 \end{pmatrix}$ unit mesh, based on two alternative models, one in which the molecular windmills are centred around an empty hollow site (see Figure S5(a)) and one in which this hollow site is occupied with an Ag adatom (Figure S5(b)).

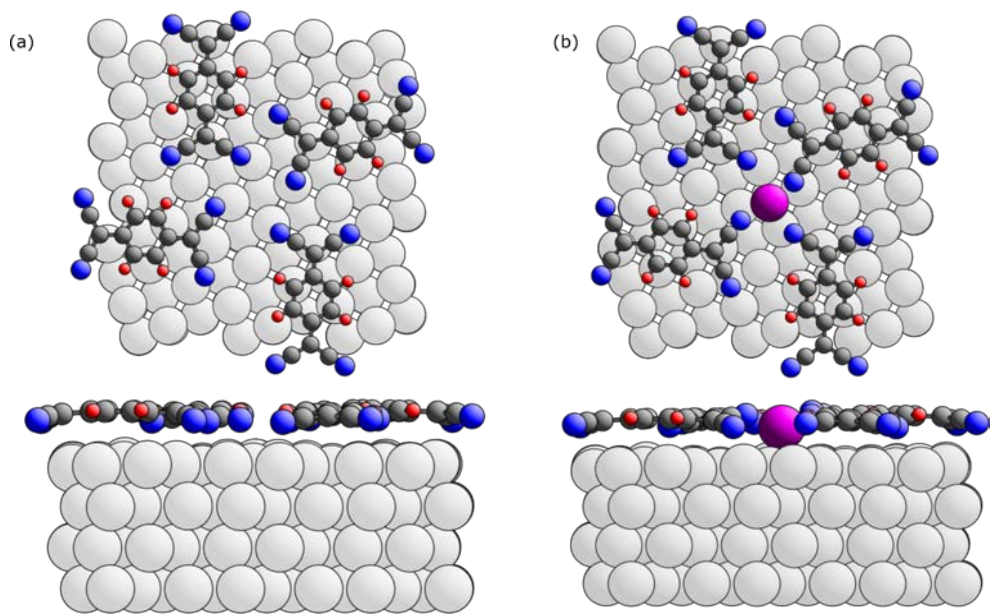

**Figure S5.** Top and side views of the optimised structures obtained from DFT calculations of two commensurate models of the local structure of the TCNQ:W2 phase (a) without Ag adatom (b) with an Ag adatom (coloured purple) in the centre of the molecular windmill based on the $\begin{pmatrix} 6 & 3 \\ -3 & 6 \end{pmatrix}$ unit mesh.

**Table S5** Adsorption energy (eV/nm$^2$) of the simulated TCNQ:W2 phase 'windmill' structure on Ag(100), with and without central Ag adatoms, computed using the two different DFT methods.

| Structural model | DFT+MBD-NL | DFT+vdW$^{surf}$ |
|---|---|---|
| TCNQ$_4$ with adatoms | 3.89 | 4.95 |
| TCNQ$_4$ – no adatoms | 3.83 | 4.90 |

Table S5 compares the adsorption energies of the two structural model. These do indicate an increased stability of the phase in the presence of an adatom, but the magnitude of the energy difference (5 or 6 meV) is very small; indeed, this is significantly less than $kT$ at room temperature (26 meV), which would imply that structures with and without Ag adatoms would be likely to be co-occupied, with similar probability, at room temperature. The failure of these calculations to provide clear support for adatom incorporation in the W2 phase is not particularly surprising. While modelling of an incommensurate phase by a commensurate phase may provide meaningful results on the substrate bonding of well-separated molecules, in a close-packed overlayer, differences in strain and consequential stress could well lead to significant differences in the energetics. Indeed, STM images showing ordered domains of the TCNQ:W2 phase also show isolated TCNQ 'windmills' (also apparently showing central Ag adatoms), as can be seen in Figure S2. This seems to indicate an apparent intrinsic lack of surface stress due to the formation of an ordered 2D array of these features, whereas this is a necessary component of the commensurate DFT-modelled structure. Nevertheless, it is notable that the TCNQ:W2 phase was only observed in coexistence with the TCNQ:LDC and TCNQ:W1 phases. Moreover, the STM images of the TCNQ:W1 phase unit mesh show that it contains four TCNQ in a 'windmill', surrounding a protrusion potentially interpreted as an Ag adatom, but also four additional TCNQ molecules with no evidence if an Ag adatom. It may be, therefore, that the energetic advantage of Ag adatom incorporation may be marginal, even after annealing.